\documentclass[preprint,showpacs,preprintnumbers,amsmath,amssymb]{revtex4}

\usepackage{graphicx}
\usepackage{dcolumn}
\usepackage{bm}

\begin{document}
\newcommand{\mf}[1]{\boldsymbol{#1}}

\title{Noisy pulses enhance temporal resolution in pump--probe spectroscopy}
\author{Kristina~Meyer}
\author{Christian~Ott}
\author{Philipp~Raith}
\author{Andreas~Kaldun}
\author{Yuhai~Jiang}
\author{Arne~Senftleben}
\author{Moritz~Kurka}
\author{Robert~Moshammer}
\author{Joachim~Ullrich}
\author{Thomas~Pfeifer}
\email{tpfeifer@mpi-hd.mpg.de}
\affiliation{Max-Planck-Institut f\"ur Kernphysik, Saupfercheckweg 1, 69117 Heidelberg, Germany}

\date{\today}

\begin{abstract}
Time-resolved measurements of quantum dynamics are based on the availability of controlled events (e.g.~pump and probe pulses) that are shorter in duration than the typical evolution time scale of the dynamical processes to be observed. Here we introduce the concept of noise-enhanced pump--probe spectroscopy, allowing the measurement of dynamics significantly shorter than the average pulse duration by exploiting randomly varying, partially coherent light fields consisting of bunched colored noise. It is shown that statistically fluctuating fields can be superior by more than a factor of 10 to frequency-stabilized fields, with important implications for time-resolved pump-probe experiments at x-ray free-electron lasers (FELs) and, in general, for measurements at the frontiers of temporal resolution (e.g.~attosecond spectroscopy). As an example application, the concept is used to explain the recent experimental observation of vibrational wave-packet motion in a deuterium molecular ion on time scales shorter than the average pulse duration. 
\end{abstract}

                             
\maketitle
Noise, the absence of order and correlation, is ubiquitous in nature. Typically, noise represents a nuisance or even a serious problem in experimental studies of structure or dynamics. With regard to spectroscopy applications, the laser has led to fast-paced progress by exhibiting remarkable coherence properties. Coherence helps to combat noise by inducing structure (fixed phase relations) in time or the spectral domain and thus enables applications such as high-resolution spectroscopy \cite{UDEM2002}, laser cooling \cite{CHU1985}, and ultrafast probing of quantum dynamics on time scales down to attoseconds \cite{HENTSCHEL2001}. Recent major achievements in several fields based on such high-coherence sources have distracted from the fact that there could be beneficial aspects of temporally noisy light sources.

One example for the benefits of noise is stochastic resonance that was found to be a far-reaching concept in natural systems~\cite{WIESENFELD1995}. It has also been recently shown that noisy light fields can enhance ionization of atoms in moderately strong laser fields~\cite{SINGH2007}. Previous work has studied the influence of noisy pulse shapes on non-resonant autocorrelation measurements~\cite{PIKE1970} and to increase spectral resolution in linear~\cite{KINROT1995} and nonlinear Coherent Raman scattering experiments~\cite{STIMSON1996, STIMSON1997, XU2008}, and two-photon absorption with quantum-correlated noise~\cite{DAYAN2004}.  However, it has not been recognized that noisy pulse shapes can improve the temporal resolution in pump--probe spectroscopy for directly measuring quantum dynamics, such as molecular or electronic wave-packet motion. It is a commonly held belief that in low-order nonlinear processes, the pulse duration limits the temporal resolution for dynamical probing experiments.
 
Here, we introduce a new concept in time-resolved spectroscopy: using varying but correlated pairs of noisy pulse shapes to enhance temporal resolution in pump-probe experiments far beyond the average-pulse-duration limit. For an example experiment discussed in this letter, the temporal resolution is increased by a factor of 10 from 30 to 3 fs. This finding thus also creates a new paradigm in the current quest for achieving the finest temporal resolution of quantum-dynamical processes: noisy light fields as an accessible alternative if dispersive compression of broadband coherent spectra is not possible.
The need to consider and rethink temporal noise has been stimulated by the development of Free-Electron Laser (FEL) Sources  as FEL pulses are known to exhibit statistically varying shapes~\cite{SALDIN2000,PFEIFER2010}. The growing field of FEL science thus provides excellent examples to demonstrate the concept. Here, we chose to describe the conceptual mechanism for the example of a sequential second-order nonlinear process, the probing of induced molecular wave-packet dynamics in a deuterium molecule. The general physical picture to understand the time-resolved pump--probe experiments with statistically varying fields is also applicable to other dynamical spectroscopy techniques. We also emphasize that the presented mechanism is universal and not even limited to optical molecular spectroscopy. It is a qualitatively new conceptual finding that can be transferred and extended to various applications in physics, technology (communications), chemistry, and the life sciences, and across various time scales and processes, wherever noisy sources of signals are involved.

\begin{figure}
\centering
\includegraphics[width=\linewidth]{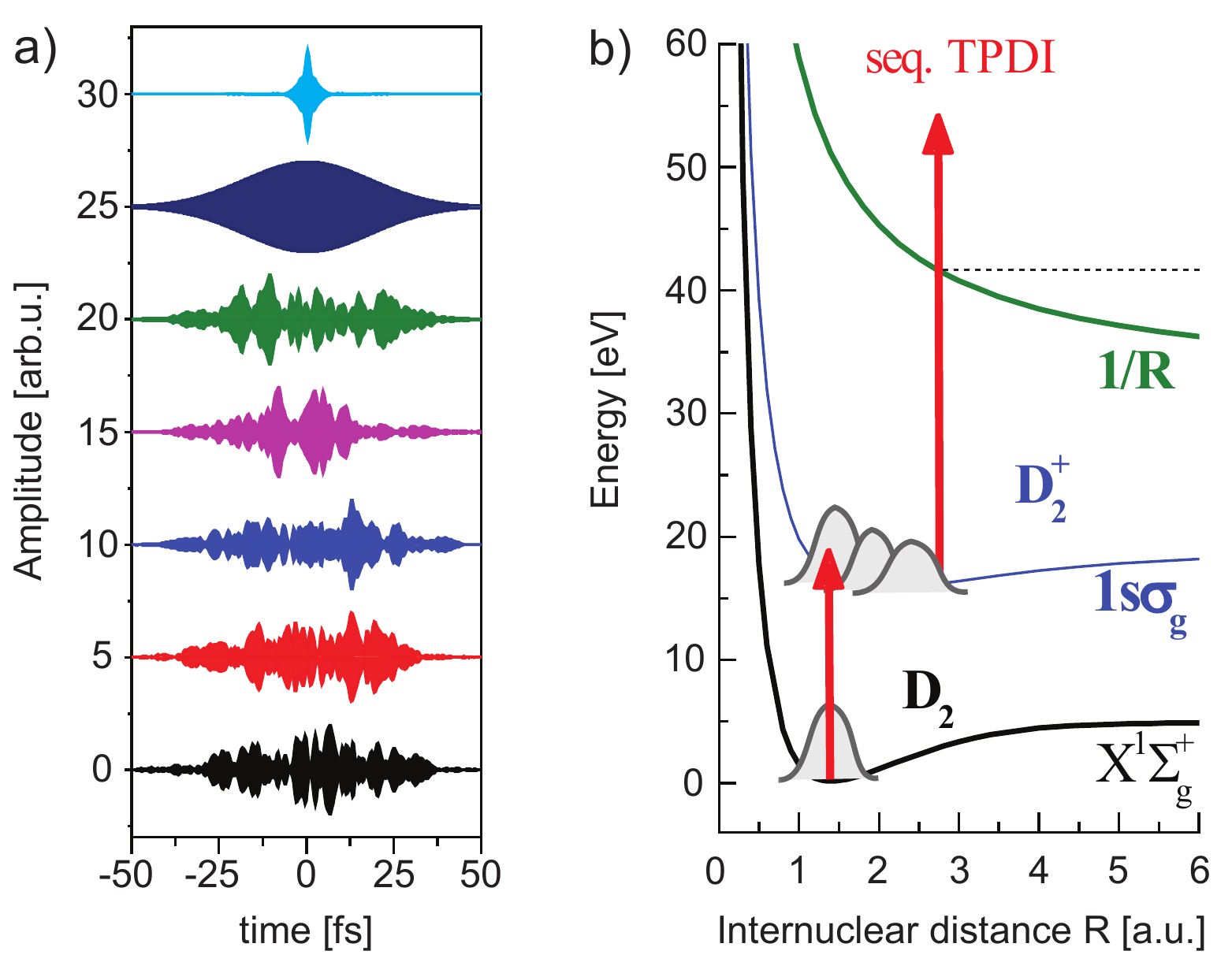}
\caption{\label{fig:Fig1}
(Color online). The essential elements of the pump--probe light--matter interaction experiment: (a) The temporal pulse shapes of (top to bottom) a 1.12~fs short pulse, a 30~fs pulse and a set of the FEL pulses. (b) A schematic of potential-energy curves of D$_{2}$ and its cations, indicating the pathway of the sequential two-photon double-ionization (TPDI) process.}
\end{figure}

The experiment discussed here was recently performed at the Free-electron LASer at Hamburg (FLASH)~\cite{JIANG2010}. This XUV-pump--XUV-probe experiment investigated the two-photon double ionization (TPDI) of D$_{2}$ using FEL pulses at an energy of 38~eV, with a bandwidth of $\sim$0.53~eV FWHM, and an average pulse duration of 30~fs.  In this study oscillations on a time scale of $\sim$20~fs were measured, which were initially unexpected given the longer average pulse duration of 30~fs.  The variation of the temporal pulse profiles from shot to shot is illustrated in Fig.~\ref{fig:Fig1}(a). 

In the experiment~\cite{JIANG2010}, the dynamics of the nuclear wave packet was monitored by measuring the kinetic energy release (KER) of the produced D$^{+}$ ions as a function of a variable time delay $\tau$ between the pump and the probe pulses.  Both pulses were derived out of the same FEL beam using a two-component split mirror (linearly cut in the center). 
The XUV-laser electric field experienced by the D$_2$ molecules in the interaction region is then the sum of the identical electric fields $E(t)$ of pump and time-delayed probe, translating into an intensity $I_\tau(t)$
\begin{equation}
I_{\tau}(t)=\left|E(t)+E(t+\tau)\right|^{2}.
\label{eq:Intensity}    
\end{equation}

In this and any other (electronic or molecular dynamics) sequential TPDI process the molecule can absorb photon 1 and photon 2 at different times $t'$ and $t''$.  As the absorption of each photon proceeds into a continuum [see Fig.~\ref{fig:Fig1}(b)] and is far from bound-state resonances (on a scale of the pulse bandwidth), a frequency-independent rate-equation model for the ionization can be used. Hereby, the probability for each individual ionization step is proportional to the intensity $I_{\tau}(t)$. In this model, the number of doubly-ionized molecules (for the example considered here) can be calculated as
\begin{equation}
N_{\mathrm{total}}(\tau)\propto\int^{\infty}_{-\infty}dt''\:\int^{t''}_{-\infty}dt'I_{\tau}(t')I_{\tau}(t'').
\label{eq:N2}   
\end{equation}

Introducing the time difference $t_{\textrm{c}}=t''-t'$ of the photon absorption and swapping integration order leads to
\begin{eqnarray}
N_{\mathrm{total}}(\tau)&\propto&\int^{\infty}_{0}dt_{\textrm{c}}\int^{\infty}_{-\infty}dt''\:I_{\tau}(t''-t_{\textrm{c}})I_{\tau}(t'') \nonumber\\
&=&\int^{\infty}_{0}dt_{\textrm{c}}\:A^{(2)}_{\textrm{c}}(t_{\textrm{c}},\tau),
\label{eq:N2final}  
\end{eqnarray}
where $A^{(2)}_{\textrm{c}}(t_{\textrm{c}}, \tau)=\int^{\infty}_{-\infty}dt\: I_{\tau}(t)\cdot I_{\tau}(t-t_{\textrm{c}})$, with $t_{\textrm{c}}$ now interpreted as correlation time, is in the following referred to as the two-dimensional autocorrelation (2dAC) function.  It will be shown below that this function is a very general object in the description of sequential second-order pump--probe experiments, irrespective of measuring molecular or electronic dynamics.

In the following, we use a molecular response function $M(E_{\textrm{KER}},t_{\textrm{c}})$ to describe the measured kinetic-energy release (KER) distribution of D$^+$ ions as a function of the temporal separation $t_{\textrm{c}}$ between two delta-like intensity pulses.  The justification for this response function derives from the following considerations: photon 1 promotes the molecule to the D$_2^+$ $1s\sigma_g$ molecular potential curve by removing one electron, while photon 2 removes the second electron to create the Coulomb-exploding D$_2^{2+}$ state.  In the time between the absorption of photon 1 and 2, a molecular wave packet evolves on the $1s\sigma_g$ state.  When the second photon is absorbed, the momentary internuclear-distance distribution of the molecular wave packet will be mapped to KER by Coulomb explosion~\cite{CHELKOWSKI1999}. The number of measured doubly ionized atoms with a specific kinetic-energy release and for a chosen time delay $\tau$ between pump and probe pulse is then given by
\begin{equation}
N(E_{\textrm{KER}},\tau)\propto\int^{\infty}_{0}dt_{\textrm{c}}\:M(E_{\textrm{KER}},t_{\textrm{c}})\;A^{(2)}_{\textrm{c}}(t_{\textrm{c}},\tau).
\label{eq:N2KER}    
\end{equation}

While the 2dAC function $A^{(2)}_{\textrm{c}}(t_{\textrm{c}}, \tau)$ contains the information on the probe light structure (including its noisy properties), the response function $M(E_{\textrm{KER}},t_{\textrm{c}})$ thus contains the physical (quantum-dynamics) information of the system to be studied. For most frequently studied sequential processes (e.g.~exciting a state and following its decoherence and decay), it could be calculated for any pump--probe scenario, also for exciting superpositions of electronic states, i.e.~time-dependent electron wave packets.  There, the response function could also depend on the observed photoelectron energies instead of or in addition to KER.

In the simulation, we used the partial-coherence method (PCM)~\cite{PFEIFER2010} to generate a set of FEL pulses starting from the average spectrum $\overline{A}(\omega)$ measured in the $D_{2}$ experiment~\cite{JIANG2010} and a random spectral phase $\phi(\omega)$. 
A Gaussian envelope as filter in the time domain accounts for the average experimental FEL pulse duration of 30~fs. 
For comparison, we consider both a bandwidth-limited Gaussian 30~fs (FWHM) pulse and a very short pulse of 1.12~fs corresponding to the Fourier transform of the average experimental spectrum, i.e.~the coherence time of the FEL pulses.  Their temporal pulse profiles are shown in Fig.~\ref{fig:Fig1}(a). 

\begin{figure}
\centering
\includegraphics[width=\linewidth]{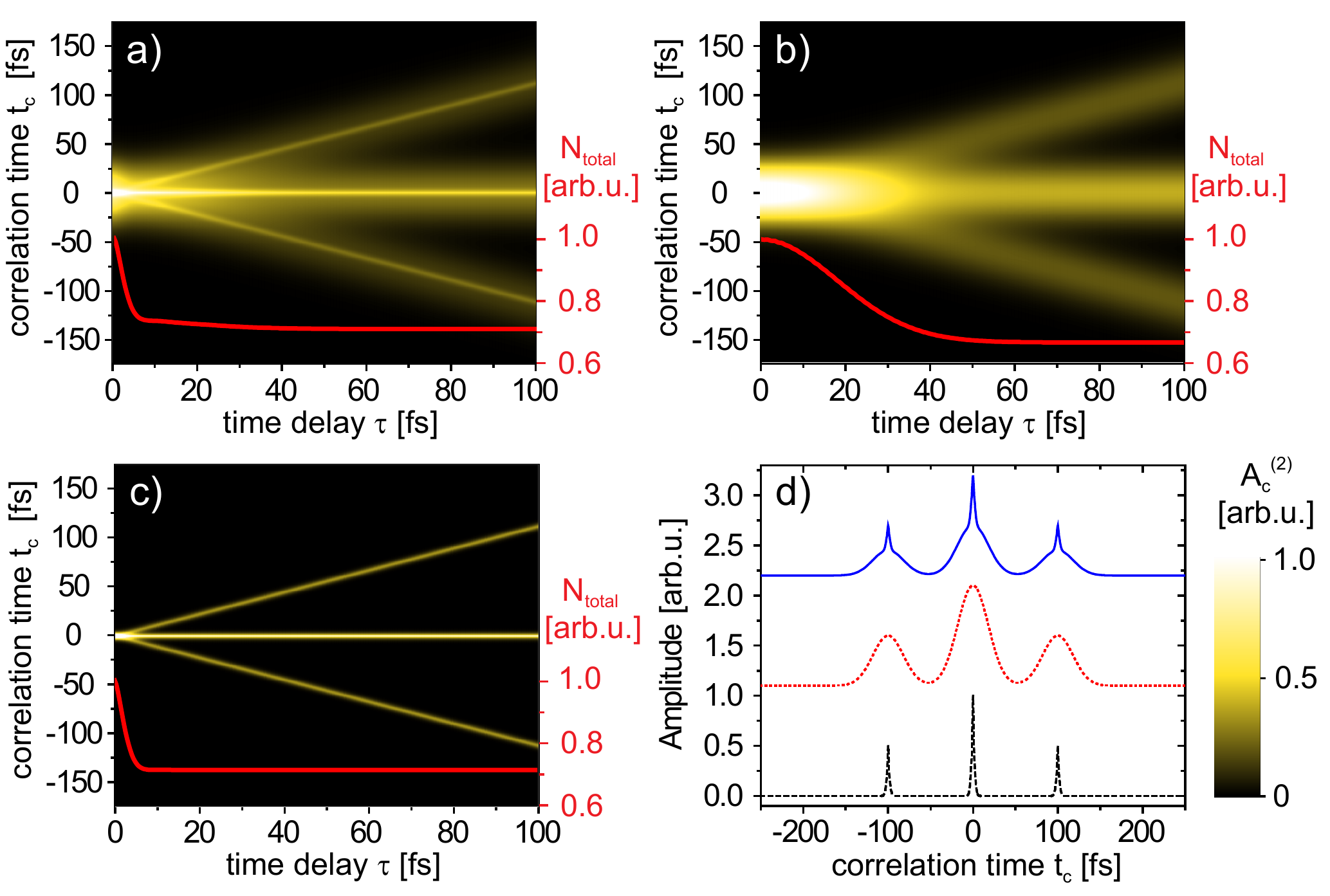}
\caption{\label{fig:Fig2}
(Color online). Two-dimensional autocorrelation functions (2dAC) $A^{(2)}_{\textrm{c}}(t_{\textrm{c}}, \tau)$, representing the intensity autocorrelation of a pair of identical (pump and probe) pulses as a function of their time delay $\tau$: (a) for a statistical average over 2000 30-fs FEL pulses, (b) for 30~fs bandwidth limited (Gaussian) pulses and (c) for 1.12~fs short pulses corresponding to the Fourier transform of the average FEL spectrum assuming a flat spectral phase. Narrow lines for $t_{\textrm{c}} = 0$ and $t_{\textrm{c}} = \pm\tau$ are observed in the averaged FEL pulse case resulting from the correlated nature of the noise in pump and probe pulses on a time scale of the coherence time. These narrow temporal features are the origin of the enhanced temporal resolution. The number of doubly ionized  molecules N$_{\textrm{total}}$ is also displayed (red curves). A lineout at a time delay of 100~fs is shown in (d) for the FEL (solid blue), the bandwidth limited 30 fs (dotted red) and the 1.12 fs short pulse (dashed black).} 
\end{figure}

The 2dAC functions $A^{(2)}_{\textrm{c}}(t_{\textrm{c}}, \tau)$ for the three different pulses are displayed in Fig.~\ref{fig:Fig2}(a)-(c) as well as the 2dAC functions integrated over $t_{\textrm{c}}$ which corresponds to the total number of doubly ionized molecules N$_{\textrm{total}}$ [cf. Eq.~({\ref{eq:N2final}})]. 
Since typically in the experiment the pump and probe beams do not overlap collinearly, optical-cycle interferences are washed out.  We account for this by convoluting the 2dAC along the $\tau$ axis with a Gaussian of width on the order of the optical cycle duration.
For the case of the bandwidth-limited 30~fs pulse, three broad lines appear, whereas these lines become very narrow for the 1.12~fs short pulse.  For averaged FEL pulses, thin lines surrounded by broader pedestals occur, which are a consequence of the correlated nature of the noise on a time scale of the coherence time. Looking at N$_{\textrm{total}}$ one can clearly see the signal enhancements for small time delays as seen before in experiments~\cite{JIANG2010A}. 
The 2dAC function for the FEL pulses closely resembles the sum of the 2dACs for the bandwidth limited 30~fs and the short 1.12~fs pulse [Fig.~\ref{fig:Fig2}(d), see also Fig.~\ref{fig:Fig4}(a)]. It is important to point out that the time resolution is determined by the time delay $\tau$ at which the main and the side peaks become separable in $t_\textrm{c}$, and also by the width of features.
As a remark, the displayed 2dAC at a given $\tau$ can also be interpreted in the following way: The main peak appearing around $t_{\textrm{c}}=0~\textrm{fs}$ represents the case that both photons are absorbed in the same pulse whereas for the smaller peaks around $t_{\textrm{c}}=\tau$ the photons originate from different pulses. For details on the number of pulses required to obtain a smooth 2dAC function as shown here, please refer to \cite{S1}.

\begin{figure}
\centering
\includegraphics[width=\linewidth]{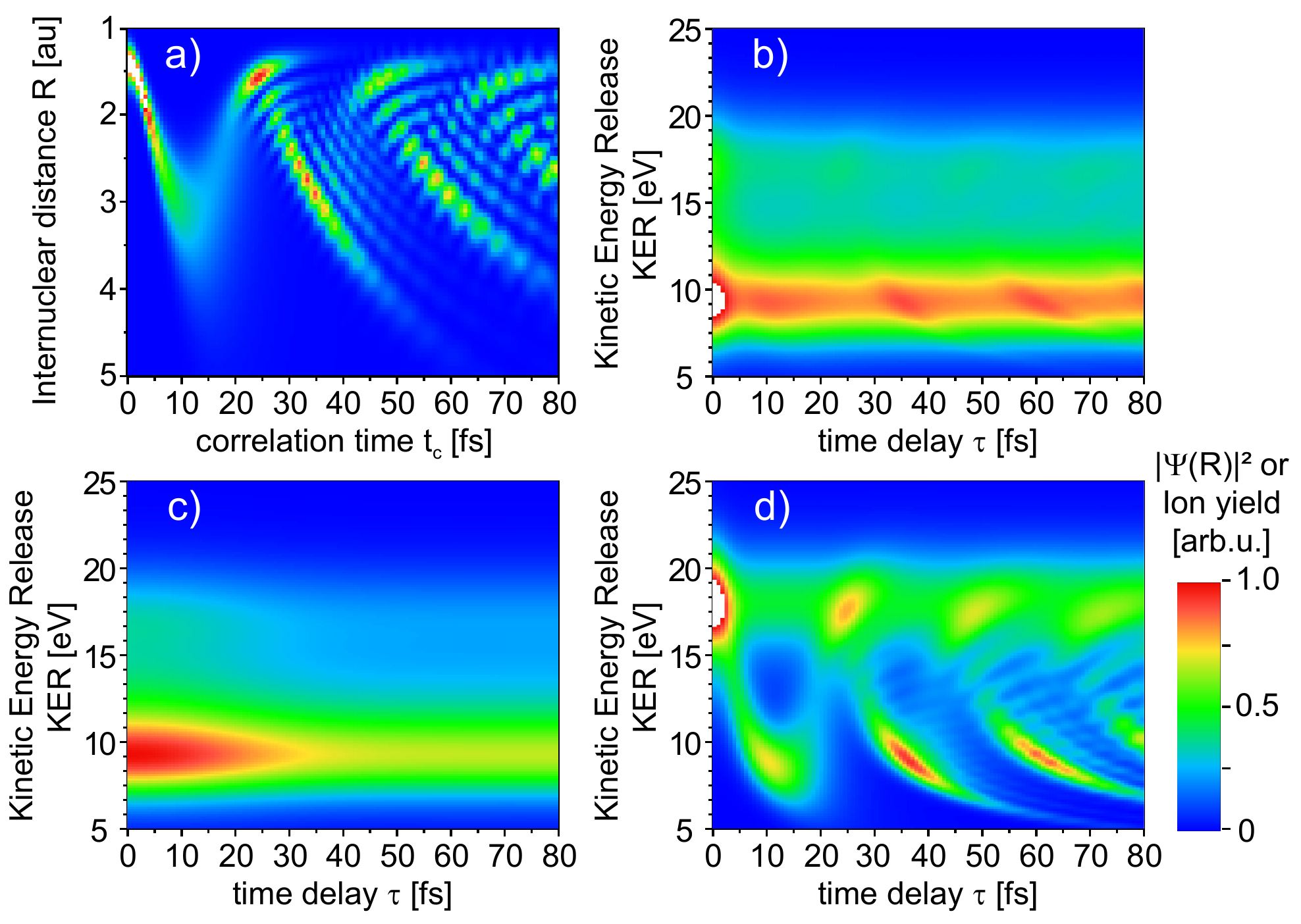}
\caption{\label{fig:Fig3}
(Color online). The internuclear wave-packet evolution $\left|\Psi(R,t_c)\right|^{2}$ is displayed in (a), yielding the molecular response function $M(E_{\textrm{KER}},t_{\textrm{c}})$ by employing the Coulomb-explosion mapping.  Using the 2dAC functions $A^{(2)}_{\textrm{c}}(t_{\textrm{c}}, \tau)$ shown in Fig.~\ref{fig:Fig2}, the KER distributions as a function of pump--probe time delay $\tau$ (as measurable in the experiment) $N(E_{\textrm{KER}},\tau)$ can be directly obtained by integration $N(E_{\textrm{KER}},\tau)\propto \int_0^{+\infty}dt_{\textrm{c}} M(E_{\textrm{KER}},t_{\textrm{c}}) A^{(2)}_{\textrm{c}}(t_{\textrm{c}}, \tau)$. Results for $N(E_{\textrm{KER}},\tau)$ are shown for an average over $2000$ FEL pulses (b), a 30~fs bandwidth-limited pulse (c) and a 1.12~fs pulse (d).  While no temporal structure is visible for the 30~fs bandwidth-limited pulse (c), the 30~fs FEL pulses, exhibiting noisy substructure, resolve the dynamics (b).}
\end{figure}

\begin{figure}
\centering
\includegraphics[width=\linewidth]{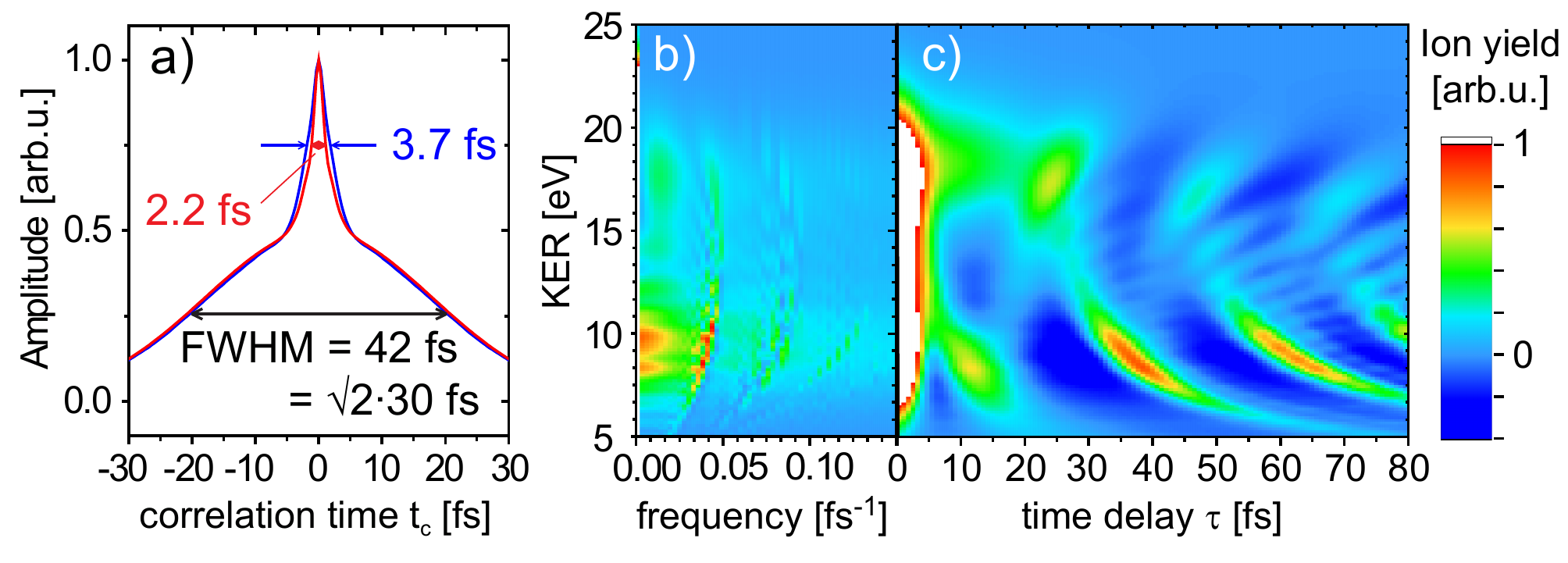}
\caption{\label{fig:Fig4}
(Color online). (a): The central section of $A^{(2)}_{\textrm{c}}$ [from Fig.~\ref{fig:Fig1}(d)] for FEL pulses (blue) is compared to the sum of 30~fs bandwidth limited and 1.12~fs short pulses (red). (b): Fourier transform (absolute value) of the averaged FEL KER-vs.-$\tau$ trace [Fig.~\ref{fig:Fig3}(b)]. (c): Pump--probe scans of the FEL pulses after removing the DC component (by Fourier transform): the result approximately recovers the dynamics obtained for the coherence-time limited 1.12~fs pulses [cf. Fig.~\ref{fig:Fig3}(d)].}
\end{figure}

The model molecular response function $M(E_{\textrm{KER}},t_{\textrm{c}})$ to illustrate the application of the 2dAC concept was chosen in the following way:  to describe the evolution of the molecular wave packet after absorption of the first photon, we calculated the time-dependent evolution of the molecular wave packet $\Psi(R,t_c)$ as a function of time $t_c$ on the 1s$\sigma_g$ curve with the initial condition of $\Psi(R,t_c=0)$ being equal to the neutral stationary molecular ground-state wave function in D$_2$.  The time-dependent Schr\"{o}dinger equation was solved using a split-step operator approach~\cite{FLECK1976,FEIT1982,PFEIFER2004A}.  The internuclear wave-packet evolution $\left|\Psi(R,t_c)\right|^{2}$ is shown in Fig.~\ref{fig:Fig3}(a), revealing an oscillatory dynamics on a time scale of approximately 20~fs.  It is then converted to the molecular response function $M(E_{\textrm{KER}},t_{\textrm{c}})$ by using the Coulomb-explosion mapping KER~$=1/R$~\cite{CHELKOWSKI1999}.

The pump--probe KER spectra $N(E_{\textrm{KER}},\tau)$ can now be calculated by using Eq.~(\ref{eq:N2KER}).  For the case of the bandwidth-limited 30~fs pulse in Fig.~\ref{fig:Fig3}(c), no dynamics can be resolved, as the $\approx$\:20~fs dynamics is washed out by the long pulse.  Employing the very short 1.12~fs pulse, Fig.~\ref{fig:Fig3}(d), the dynamics is clearly resolved.  Interestingly, although the average pulse duration of the FEL is also 30 fs, the dynamics can be retrieved [see Fig.~\ref{fig:Fig3}(b)] even for averaging over many differently structured FEL pulses. We obtain as oscillation period 23~$\pm$~1~fs~\cite{S2} which is in excellent agreement with the experimental result of 22~$\pm$~4~fs~\cite{JIANG2010}. 
The contribution of signal at higher KER is underestimated in our simulation for two reasons: (1) The cross-section for the ionization step from D$_2^+$ to D$_2^{2+}$ is assumed to be independent of the internuclear distance. (2) The direct (nonresonant) TPDI process is neglected which would lead to a contribution at higher KER.

It should be pointed out that the sum of the 2dACs for 1.12~fs short and 30~fs bandwidth-limited pulses is showing slightly finer "spikes" as the FEL average 2dAC, as can be seen in Fig.~\ref{fig:Fig4}(a) for the case $\tau=100~\textrm{fs}$.  The obtained temporal resolution of the dynamics will thus always be slightly reduced for the case of statistically fluctuating FEL pulses, compared to the case of extremely short coherence-time duration laser pulses, but still by a factor of 10 better than the 30~fs average pulse duration [see Fig.~\ref{fig:Fig4}(a)]. 
It is particularly interesting that the here-discussed phenomenon of noise-enhanced resolution is independent of the average pulse duration.  Short time-scale spikes will always be present in the 2dACs as long as the coherence time of the source is sufficiently short (the average spectrum is sufficiently broad), just the ratio of spike-to-pedestal area would decrease and thus lower the usable dynamical signal vs.~the background from the long pulse.

The fast modulations arising by the wave-packet motion can be uncovered by means of a Fourier transform along the time delay $\tau$ axis [see Fig.~\ref{fig:Fig4}(b)], removing the DC contribution and then transforming back into the time-delay domain [see Fig.~\ref{fig:Fig4}(c)]. The dynamics and time-dependent shape of the molecular wave packet can thus be recovered with very high accuracy, closely resembling the expected signal for the extremely short (coherence-time limited) pulse shown in Fig.~\ref{fig:Fig3}(d). Further detail information on how counting statistics in the experiment influences the resolution of this method can be found in~\cite{S1}.

In conclusion, we presented the concept of noise-enhanced temporal resolution in pump--probe experiments.
We thus also provide the physical mechanism behind the recent surprising observation of sub-pulse-duration dynamics in D$_2$ measured with statistically varying FEL pulses. We find that dynamical features much shorter than the average pulse duration (on the order of the coherence time of the individual pulses) can be resolved by employing the correlated nature of the noise in the pump and the probe pulses. Importantly, it should be noted that the choice of the molecular response function -- also neglecting excitations to all other repulsive potential energy curves that do not result in oscillatory wave-packet dynamics -- in this work was only to demonstrate the mechanism and the applicability of the 2dAC functions for well defined and statistical pulse shapes. While a different and better-suited response function can be used for the described example system of D$_2^+$, the applicability of the presented mechanism is more general:
Important consequences arise in the current race for shorter and shorter pulses and better dynamical resolutions in attosecond (and beyond) science. The mechanism could help for the case of extremely broadband high-harmonic spectra that are currently generated in experiments~\cite{SERES2005, MASHIKO2009, POPMINTCHEV2009, CHEN2010} and that can likely not be compressed to their bandwidth-limited few-as duration due to the absence of suitable dispersion-compensating optics. The herein presented results and novel paradigm suggest a new route: if compression is not possible, it is sufficient to statistically vary the spectral phase of the pulses. In~\cite{S3} we show in detail the example for the attosecond pulse case, including the viability of the method for not fully incoherent spectral phases. This finding also has important consequences for nonlinear spectroscopy in the life sciences, as it enables time-resolved probing through moving quasi-randomly dispersive and scattering soft tissue, and could have other important applications in communications science and biological signal transmission to increase robustness of data transmission in the presence of noise.

We gratefully acknowledge funding from the Max-Planck Gesellschaft within the scope of the Max-Planck Research Group (MPRG) program.

\bibliographystyle{apsrev}

\end{document}


\newcommand{\mf}[1]{\boldsymbol{#1}}

\title{Supplemental material to "Noisy pulses enhance temporal resolution in pump--probe spectroscopy"}
\author{Kristina~Meyer}
\author{Christian~Ott}
\author{Philipp~Raith}
\author{Andreas~Kaldun}
\author{Yuhai~Jiang}
\author{Arne~Senftleben}
\author{Moritz~Kurka}
\author{Robert~Moshammer}
\author{Joachim~Ullrich}
\author{Thomas~Pfeifer}
\email{tpfeifer@mpi-hd.mpg.de}
\affiliation{Max-Planck-Institut f\"ur Kernphysik, Saupfercheckweg 1, 69117 Heidelberg, Germany}

\maketitle

\section{Comparison to the experiment and the influence of statistics}

\begin{figure}[b]
\centering
\includegraphics[width=0.5\linewidth]{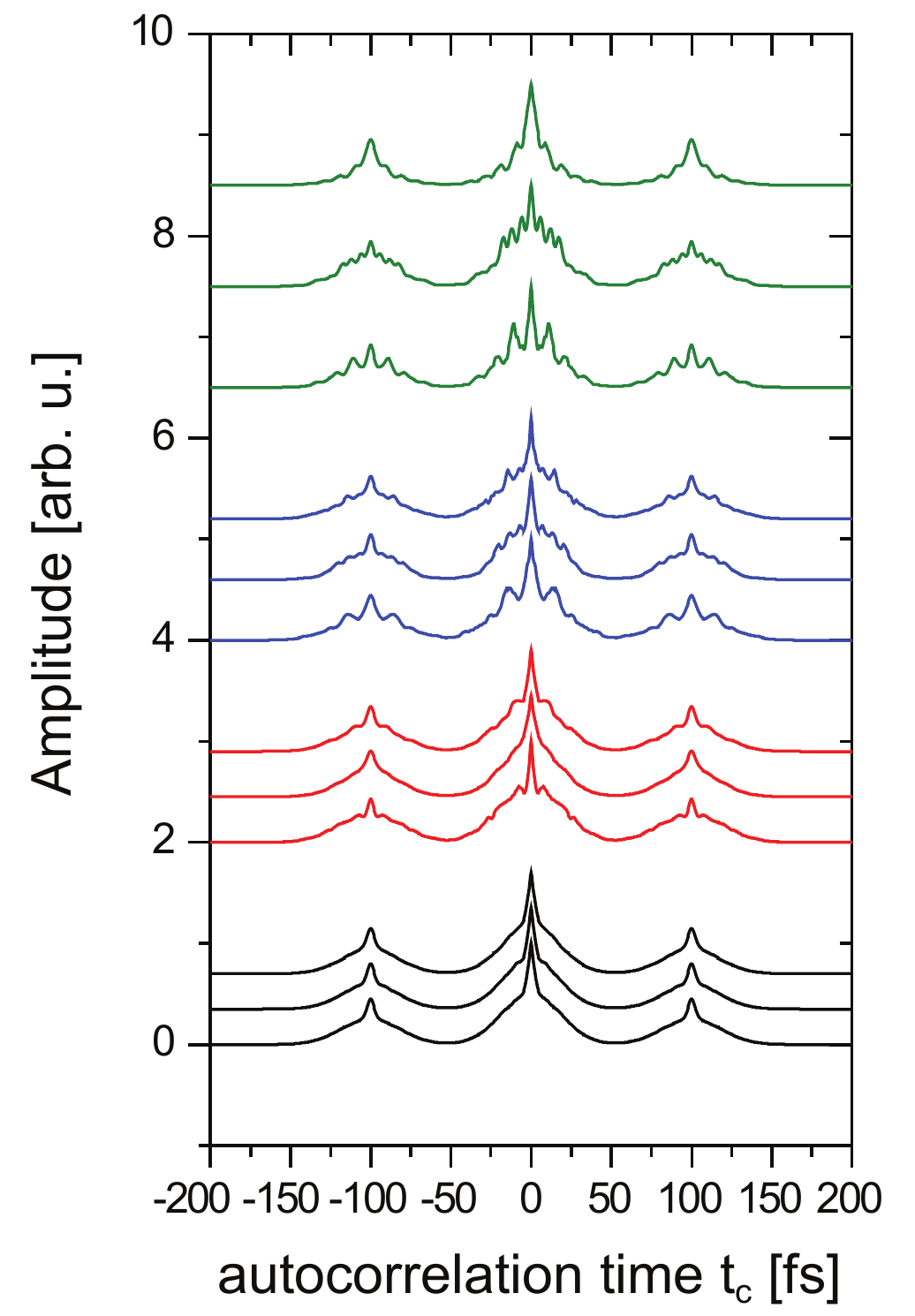}
\caption{\label{fig:FigS1}
Lineouts of the 2dAC function $A^{(2)}_{\textrm{c}}(t_{\textrm{c}}, \tau)$ at $\tau = 100$ fs for different numbers of averaged pulses, three different averages each: single FEL pulses (green), average over three (blue), ten (red) and 100 pulses (black).  For single random pulses several spikes occur at random $t_c$, depending on the individual pulse shape.  The larger the number of averaged pulses the more the (2d-)autocorrelation function converges to its final shape: the singular coherence spike on top of the broad pedestals, which results in the increased temporal resolution.  Thus, when averaging over only few pulse shapes per time delay in a measurement, the temporal resolution would be less well determined.}
\end{figure}

\begin{figure}[b]
\centering
\includegraphics[width=0.5\linewidth]{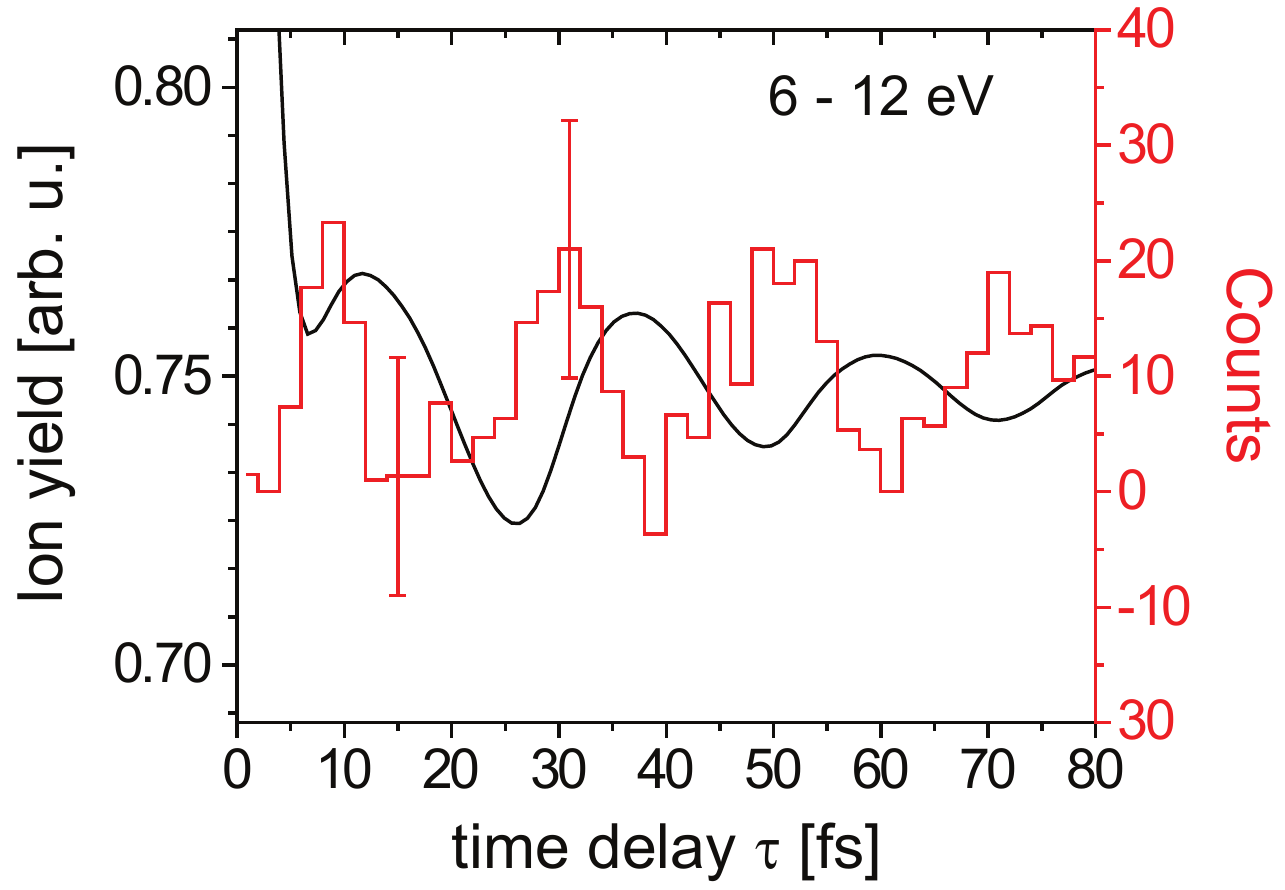}
\caption{\label{fig:FigS2}
Ion yields depending on the pump--probe time delay for the KER regime 6 - 12 eV. The experimental curves (red) taken from~\cite{JIANG2010} and the ion yields obtained from our simulation (black) are displayed.  The results agree except for a temporal shift, which often is an unknown parameter in the experiment and needs to be obtained by fitting to a model.  The numerical oscillation period of 23 $\pm$ 1~fs (obtained by statistics over the peak-to-peak and valley-to-valley delay times) agrees with the experimental period of 22 $\pm$ 4~fs.}
\end{figure}

In our simulation we averaged over 2000 pulses for the results shown in the main text of our letter.  Before quantifying the impact of statistics on the temporal resolution it should be noted that the noise-enhanced resolution as discussed here should not be confused with other (e.g.~spatial superresolution/deconvolution) techniques where the centroid of an intensity distribution can be determined much more accurately ($\propto 1/\sqrt{N}$, $N$ the number of measurements) than its width.  In our case, the resolution is defined by the sharp temporal feature in the 2d-autocorrelation functions, which is, however, superimposed on a broader component (resulting in a "background" in the measurement).  We show the evolution of this sharp component for averaging over different numbers of pulses in Fig.~\ref{fig:FigS1}. For single FEL pulses, the 2dAC function varies significantly from pulse to pulse (green curves). A spike is always located at the top of the broad pedestals ($t_c=\tau$), but there are several other spikes whose positions change from pulse to pulse due to the random pulse shapes. If we start averaging over several $N$ FEL pulses, the different peaks are washed out more and more with increasing $N$ except for the narrow spikes on top of the pedestals. Our temporal resolution is given by the locations, intensities, and widths of these spikes, becoming maximal if only the singular spike at $t_c=\pm\tau$ is obtained for averaging over a large-enough number $N$ of pulses.  For the case considered here, $N=100$ seems to be sufficient to obtain maximum temporal resolution.  This number will vary, however, with the parameters average pulse duration and average spectrum (i.e.~coherence time), and needs to be obtained for each given set of parameters, e.g.~by performing a convergence analysis as described here.

In order to quantitatively compare our simulation to the experiment the measured ion yield is displayed in Fig.~\ref{fig:FigS2} as well as the expected yield derived from our simulation. Thereby, we summed over the KER range of 6 - 12 eV (cf. Fig.~3(b) of the main text) and determined the period of the oscillation to be 23 $\pm$ 1~fs. This is in excellent agreement with the experimentally obtained period of 22 $\pm$ 4~fs~\cite{JIANG2010}. Simulation (black curve) and experiment (red curve) agree well except for a shift along the axis of the time delay $\tau$. However, the position of zero time delay in the experiment is often unknown and needs to be matched to the model. 

\begin{figure}[b]
\centering
\includegraphics[width=0.95\linewidth]{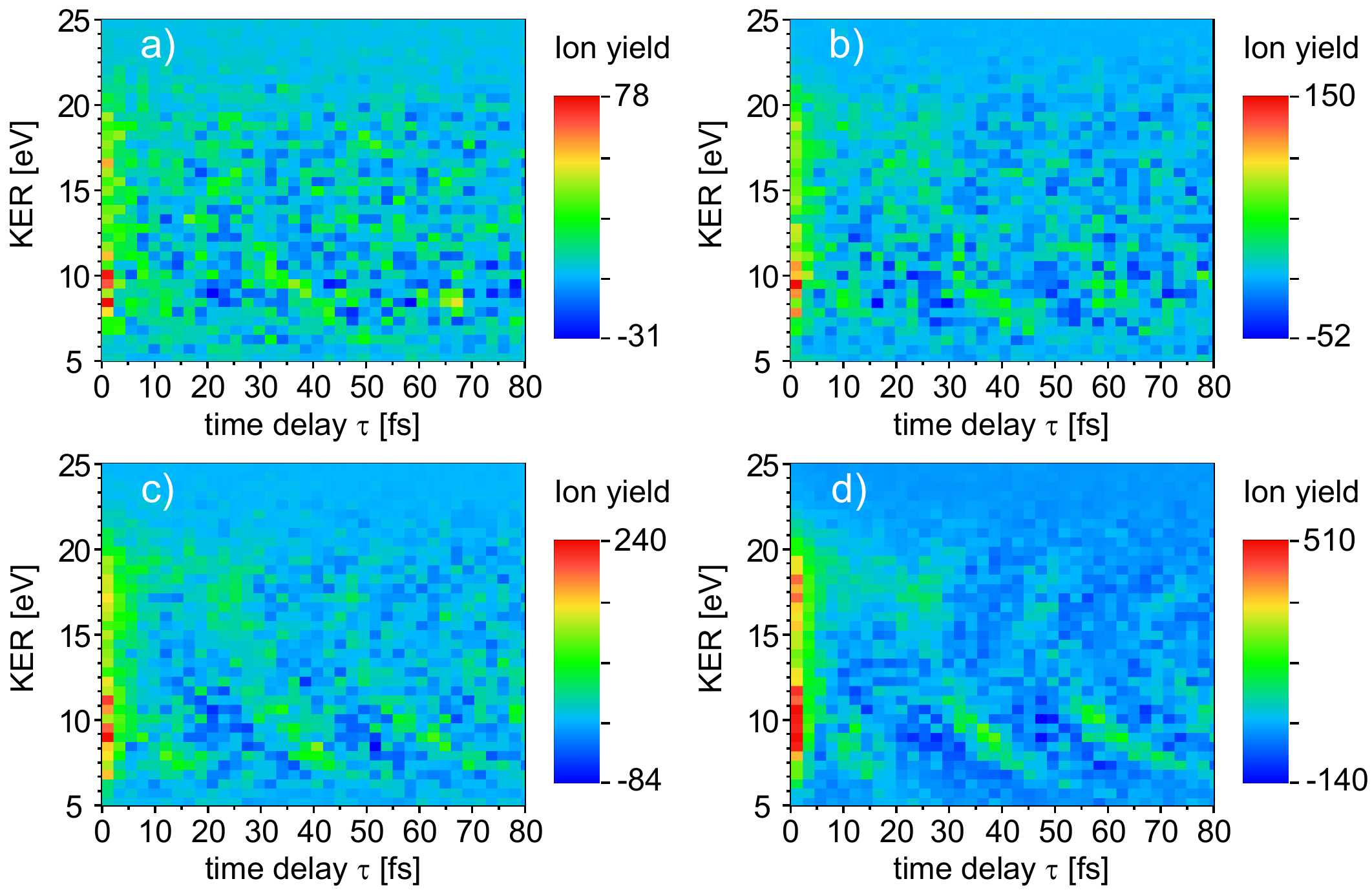}
\caption{\label{fig:FigS3}
Simulation of Fourier filtered KER-vs.$\tau$ traces for different total numbers of counts (Poissonian statistics): (a) 4.5$\cdot$10$^5$, (b) 9$\cdot$10$^5$, (c) 1.8$\cdot$10$^6$ and (d) 4.5$\cdot$10$^6$ counts.}
\end{figure}

In addition, we examined the influence of counting statistics in the experiment, in particular for the reconstruction of the wave-packet motion by Fourier transform as described in the main text.  Here, we simulated four different cases of experimental random Poissonian count statistics resulting in a total number of counts of 4.5$\cdot$10$^5$, 9$\cdot$10$^5$, 1.8$\cdot$10$^6$ and 4.5$\cdot$10$^6$, and studied their impact on the Fourier-filtered KER-vs.-$\tau$ trace (see Fig.~\ref{fig:FigS3}), shown in the main paper in Fig.~4(c).  For low count numbers, the dynamics in the lower KER range is difficult to identify [Fig.~\ref{fig:FigS3}(a)].  The wave-packet-dynamics structures become more and more apparent with increasing number of total counts.  For about 4.5$\cdot$10$^6$ counts [Fig.~\ref{fig:FigS3}(d)], the dynamics can be clearly recovered.

\section{General applicability, example for attosecond pulses}

In order to further illustrate that the introduced concept is not only applicable to FEL pump--probe experiments, but more general as we stated in the last paragraph of the paper, we here apply our method to broadband high-harmonic attosecond-pulsed light.  As an example, we consider pump--probe experiments using noisy attosecond pulses.  Attosecond pump--probe experiments will have a wide-spread applicability throughout the physical and chemical sciences and a proof-of-principle experiment has very recently been published~\cite{TZALLAS2011}.  As attosecond pulses are typically coherent but exhibit a chirp which is not easily compressed by optical elements, it is interesting to study the possibility of using our noise-enhancement method to realize effectively bandwidth-limited resolution also in the attosecond pulse case.  Here, it is important to study the influence of residual coherence existing in the attosecond pulse.  Maximum coherence would be obtained if all important laser parameters such as intensity, carrier-envelope phase (CEP) and duration can be kept constant from shot to shot, a case which is difficult to realize in experiments.

\begin{figure}
\centering
\includegraphics[width=0.5\linewidth]{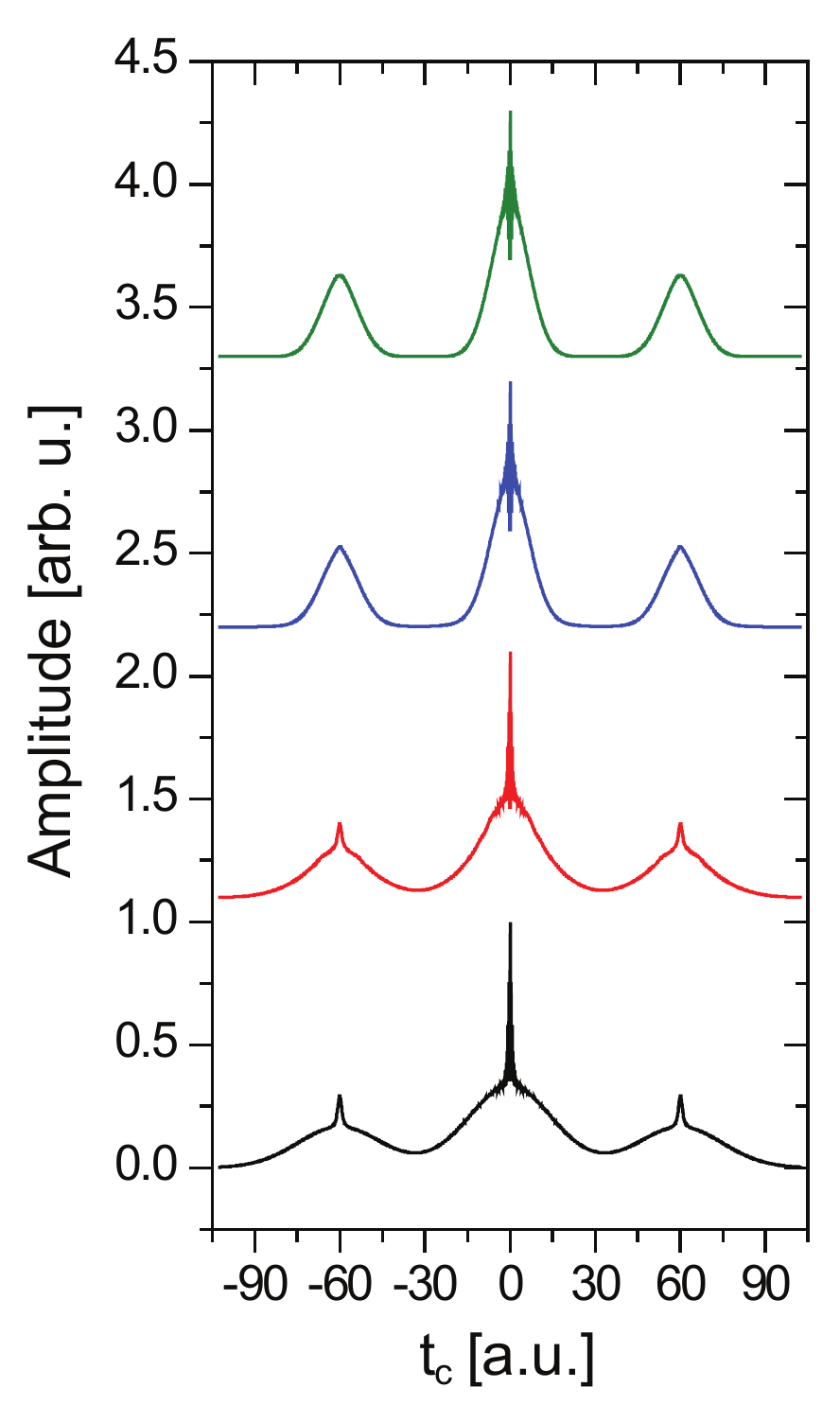}
\caption{\label{fig:FigS4}
Influence of residual spectral coherence and increasing spectral phase noise for the attosecond pump--probe case.  Lineouts of the 2dAC function $A^{(2)}_{\textrm{c}}(t_{\textrm{c}}, \tau)$ at $\tau = 600$ a.u.~for different amounts of spectral phase noise in a range of $x$ added to the original attosecond-pulse spectral phase: $x = 0.2\pi$ (green), $0.6\pi$ (blue), $1.4\pi$ (red), and $2\pi$ (black).}
\end{figure}

For the calculation we use an isolated (gated) attosecond pulse generated by an IR light pulse with a center wavelength of 800~nm, a pulse duration of 5~fs FWHM and a peak electric field of 0.2~a.u. The attosecond pulse exhibits a chirped spectrum with a pulse duration of about 9 a.u. ($\approx$\:220 as) FWHM. The calculation of the two-dimensional autocorrelation (2dAC) function $A^{(2)}_{\textrm{c}}(t_{\textrm{c}}, \tau)$ was performed as described in the paper. Firstly, we add noise to the attosecond pulse original spectral phase, numerically discretized in frequency steps of $\Delta \omega=0.2\;\textrm{fs}^{-1}$. In order to study the influence of the amount of added phase noise, the phase was randomly varied at each frequency with maximum random excursions $x$ between zero and $2\pi$ from its original value.  Secondly, an average pulse duration of 600~as was assumed by a Gaussian envelope as filter in the time domain, in order to model the maximum temporal noise introduced, e.g.~by reflecting of rough mirrors for an experimental implementation of the noise source.  The lineouts of the 2dAC functions at $\tau = 60$~a.u. for the different values of $x$ are shown in Fig.~\ref{fig:FigS4}. As in the example of FEL pulses, but now starting from a chirped coherent pulse, spikes on top of broader pedestals occur, leading to an enhanced temporal resolution.  It can be clearly seen that the result depends on the amount of noise imprinted on the spectral phase. A minimum critical amount of noise is required to achieve enhanced temporal resolution. In our simulation, this occurs near a critical $x_c$ = 1.4$\pi$. For lower values of $x$, the spikes at the side peaks at $t_\textrm{c}= \pm 60$ a.u.~disappear. For $x$ = 0.6$\pi$, the remaining pedestal does not show a Gaussian, but a rather triangular shape, representing the transition range to the coherent case. For very low $x$, i.e.~0.2$\pi$, the attosecond pulses can be regarded as being nearly fully coherent and the 2dAC lineout is determined by the original attosecond pulse duration ($\approx$\:220~as) and shape. The width of the pedestals for higher $x$ corresponds to the average statistical pulse duration of 600~as. 

The specific features of the 2dAC function shown above---the very narrow spikes with a FWHM of approximately 1.5 a.u.\:($\approx$ 36~as thus corresponding to an effective pulse duration of 25~as)---thus provide substantially (order of magnitude) enhanced temporal resolution over the original 220~as pulse duration. The obtained temporal resolution nearly reaches the bandwidth limit (i.e.~assuming flat spectral phase of original attosecond pulse spectrum and no additional temporal filter) of about 0.8~a.u.\:($\approx$ 19~as) pulse duration, which is experimentally not feasible due to the lack of compensating optics for dispersion correction.  The noisy-pulse scheme with the attainable atomic-unit time resolution would thus be applicable to measurements of electron dynamics based on sequential (i.e.~resonant, ionizing) processes, such as creating (multi-)electronic wave-packet motion in (ionic) excited states.  The herein presented calculation shows that the concept of noise-enhanced temporal resolution is not only restricted to experiments with FEL pulses, but also applicable to attosecond pump--probe spectroscopy, paving the way to observing electron dynamics on the few attosecond time scale, especially when high-intensity long-wavelength drivers are able to push high-harmonic cutoffs to several keV and beyond.

As a note towards an experimental realization, the noise could be introduced for instance using non-stationary mirrors with rough surfaces or, possibly more efficiently, employing multilayer-nanofluidic mirrors where different spectral parts of the attosecond pulse are randomly phased or delayed at different penetration depths into such a mirror.  It should be noted that the noise-enhanced method is  even superior to creating a perfect multilayer mirror for phase compensation (which seems unlikely for future coherent soft-x-ray ranges exceeding several keV), due to the sensitivity of the attosecond pulse chirp on pulse intensity and CEP fluctuations.  A noise-enhanced scheme is inherently tolerant to (and even augmented by) such chirps and thus phase variations of the attosecond pulses and robustly provides high temporal resolution.

\bibliographystyle{apsrev}